\documentclass[sigconf,screen,nonacm]{acmart}

\usepackage[inline]{enumitem}

\AtBeginDocument{%
  \providecommand\BibTeX{{%
    \normalfont B\kern-0.5em{\scshape i\kern-0.25em b}\kern-0.8em\TeX}}}

\setcopyright{acmlicensed}
\copyrightyear{2018}
\acmYear{2018}
\acmDOI{XXXXXXX.XXXXXXX}

\acmJournal{JACM}
\acmVolume{37}
\acmNumber{4}
\acmArticle{111}
\acmMonth{8}

\acmISBN{978-1-4503-XXXX-X/18/06}

\begin{document}

\title{Unexplored Frontiers: A Review of Empirical Studies of Exploratory Search}


\author{Alan Medlar}
\affiliation{%
  \institution{Department of Computer Science \\
University of Helsinki}
  \city{Helsinki}
  \country{Finland}}
\email{alan.j.medlar@helsinki.fi}

\author{Denis Kotkov}
\affiliation{%
  \institution{Department of Computer Science \\
University of Helsinki}
  \city{Helsinki}
  \country{Finland}}
\email{denis.kotkov@helsinki.fi}

\author{Dorota G\l owacka}
\affiliation{%
  \institution{Department of Computer Science \\
University of Helsinki}
  \city{Helsinki}
  \country{Finland}}
\email{dorota.glowacka@helsinki.fi}
\renewcommand{\shortauthors}{Medlar, Kotkov and G\l owacka}

\begin{abstract}
This article reviews how empirical research of exploratory search is conducted. We investigated aspects of interdisciplinarity, study settings and evaluation methodologies from a systematically selected sample of 231 publications from 2010-2021, including a total of 172 articles with empirical studies. Our results show that exploratory search is highly interdisciplinary, with the most frequently occurring publication venues including high impact venues in information science, information systems and human-computer interaction. However, taken in aggregate, the breadth of study settings investigated was limited. We found that a majority of studies (77\%) focused on evaluating novel retrieval systems as opposed to investigating users' search processes. Furthermore, a disproportionate number of studies were based on scientific literature search (20.7\%), a majority of which only considered searching for Computer Science articles. Study participants were generally from convenience samples, with 75\% of studies composed exclusively of students and other academics. The methodologies used for evaluation were mostly quantitative, but lacked consistency between studies and validated questionnaires were rarely used. In discussion, we offer a critical analysis of our findings and suggest potential improvements for future exploratory search studies.

\end{abstract}

\ccsdesc[500]{Information systems~Information retrieval}
\ccsdesc[500]{Human-centered computing~User studies}

\keywords{exploratory search, information retrieval, systematic review}


\maketitle

\section{Introduction}

Exploratory search was conceptualised as the interdisciplinary study of how search could be used to facilitate complex tasks related to learning and investigation \cite{marchionini2006exploratory}. These search tasks are open-ended, give rise to evolving information needs and are highly subjective in terms of the relevance of search results to those needs \cite{white2009exploratory, athukorala2016exploratory}. Research in exploratory search, therefore, necessitates empirical studies that blend aspects of information science, information retrieval and human-computer interaction to build retrieval systems that support exploratory search activities and to understand related search behaviors. While the complexity of exploratory search has led to definitions that are broad in scope \cite{white2009exploratory}, empirical studies tend to define exploratory search indirectly according to the attributes of search tasks believed to elicit exploratory behavior \cite{wildemuth2012assigning, li2010exploration}. 

In this article, we present a systematic review of empirical studies of exploratory search. The motivations for this review are twofold. 
First, since White and Roth there have been no reviews of exploratory search that surveyed the field as a whole \cite{white2009exploratory}. Furthermore, while there have been reviews that focused on specific aspects of exploratory search, such as search tasks \cite{wildemuth2012assigning} and information seeking models \cite{palagi2017survey}, there have been no systematic reviews of the research literature. The period considered in our review, 2010-2021, was chosen to follow on from White and Roth. Second, we believe that analyzing empirical studies will help to clarify what the research community believes exploratory search is in practice or, at the very least, what aspects of exploratory search receive more attention than others. Furthermore, the related work sections of articles only give a limited snapshot of current research, whereas systematic reviews can provide an unbiased view of a given field of study.

This article aims to provide as complete a picture of research in exploratory search as possible. We look at the interdisciplinary nature of exploratory search, identifying the most common publication venues and how authors from different disciplines position their research contributions. We enumerate what is studied in empirical studies, covering retrieval systems, search domains and simulated work tasks. We additionally investigate the experimental design characteristics of exploratory search studies, including the types of participants studied and what data is collected during evaluation. Lastly, we provide a discussion where we put our findings into context. We detail what we see as problems associated with current research practices and 
suggest potential improvements for future exploratory search studies

\section{Background}

Marchionini's original conception of exploratory search was of a search process that blends the use of queries with browsing strategies to satisfy complex information needs \cite{marchionini2006exploratory}. To distinguish exploratory search from other kinds of information search, Marchionini describes three types of search activity: lookup, learn and investigate. Lookup search (or known item search) requires the formulation of well-defined, unambiguous search queries to retrieve discrete, well-structured facts or specific documents. Learn and investigate, however, are seen as core activities in exploratory search, necessitating the use of browsing strategies, multiple search iterations and the assessment of many potentially relevant search results \cite{athukorala2016exploratory, athukorala2016beyond,mitsui2018much,kotzyba2017exploration,palmquist2000cognitive}. For example, searchers engaged in learning tasks, such as self-directed learning, need to evaluate information from various sources (websites, scientific literature, videos, etc.) for the purpose of knowledge acquisition. Whereas investigation tasks involve information synthesis, discovery and planning, e.g.,~finding relevant articles for a scientific literature review \cite{medlar2016pulp}, which is more reliant on recall (i.e.,~maximising the number of relevant documents identified). 

In contrast, White and Roth define exploratory search as both an information seeking problem context and a search process \cite{white2009exploratory}. As a problem context, users engaged in exploratory search are motivated by either a need to address a knowledge gap \cite{belkin1982ask} or satisfy their curiosity \cite{pace2004grounded}. As a search process, exploratory search involves significant cognitive processing and interpretation, and is, therefore, distinct from traditional information retrieval scenarios \cite{marchionini2006exploratory}. White and Roth emphasise the role of uncertainty in exploratory search. Users engaged in exploratory search may be unfamiliar with their search domain, be unsure how to achieve their goals or be unsure about what their search goals are. White and Roth carry the theme of uncertainty through to a model of exploratory search behavior composed of two phases: exploratory browsing and focused searching \cite{white2009exploratory}. Exploratory browsing is necessary due to the uncertainty of the initial problem context, whereas focused searching results from decreasing uncertainty as users learn about the search domain. However, others have suggested that these distinctions are unnecessary and exploratory search can be sufficiently described in terms of other information seeking models \cite{savolainen2018berrypicking}.

Wildemuth and Freund highlight how experimental studies tend to define exploratory search indirectly in terms of cognitive and behavioral search task attributes \cite{wildemuth2012assigning}. 
Cognitive attributes of exploratory search tasks include:
\begin{enumerate*}[label=(\roman*)]
    \item learning and investigation related search goals \cite{marchionini2006exploratory},
    \item conceptually broad or underspecified search task descriptions \cite{kules2008creating},
    \item uncertainty with respect to the clarity of the information need, the topic, the available documents or search outcomes \cite{white2009exploratory}, 
    \item ill-structured underlying questions, problems or information needs \cite{macmullin1984problem}, 
    \item dynamic, in terms of having evolving information needs \cite{bates1989design}, 
    \item multi-faceted, with search outcomes covering multiple concepts (e.g.~\cite{toms2007task}),
    \item ``not too easy'', in terms of the inherent task difficulty \cite{kules2008creating} 
        or procedural complexity, e.g.~involving many subtasks \cite{diriye2010revisiting}, and
    \item accompanied by cognition, such as decision-making or sensemaking \cite{dervin1998sense}.
\end{enumerate*}
Whereas, from a behavioral perspective, exploratory search tasks can feature:
\begin{enumerate*}[label=(\roman*)]
    \item open-ended problems for which there is no specific answers \cite{marchionini1993information},
    \item multiple target items resulting from open-ended learning tasks \cite{marchionini2006exploratory}, and 
    \item occur over time and, therefore, involve multiple queries and search sessions \cite{white2009exploratory}.
\end{enumerate*}
Our review complements Wildemuth and Freund's analysis by providing an overview of the field to understand what is studied in exploratory search and how it is evaluated experimentally.



In summary, 
exploratory search is broadly defined with experimental studies describing it in terms of numerous cognitive and behavioral attributes.
In our meta-analysis, we investigate 
bibliographic data (e.g.,~publication venue, keywords),
what is studied (retrieval systems, search domains, work tasks), and
how evaluation is conducted (experimental design, choice of participants, data collection).
Our goal is to understand how investigators present exploratory search in practice, expose underexplored areas of the problem space and identify future research challenges.

\section{Review Procedure}

We conducted our review using an analytical approach based on a representative sample of published work spanning a diverse range of publication venues. The selection of publications was based on the QUOROM statement \cite{moher2000improving}, that defines a procedure for conducting meta-analyses. 
Figure 1 provides an overview of how papers were identified for inclusion in the study.

\begin{figure*}[t]
\centering
  \includegraphics[width=0.7\linewidth]{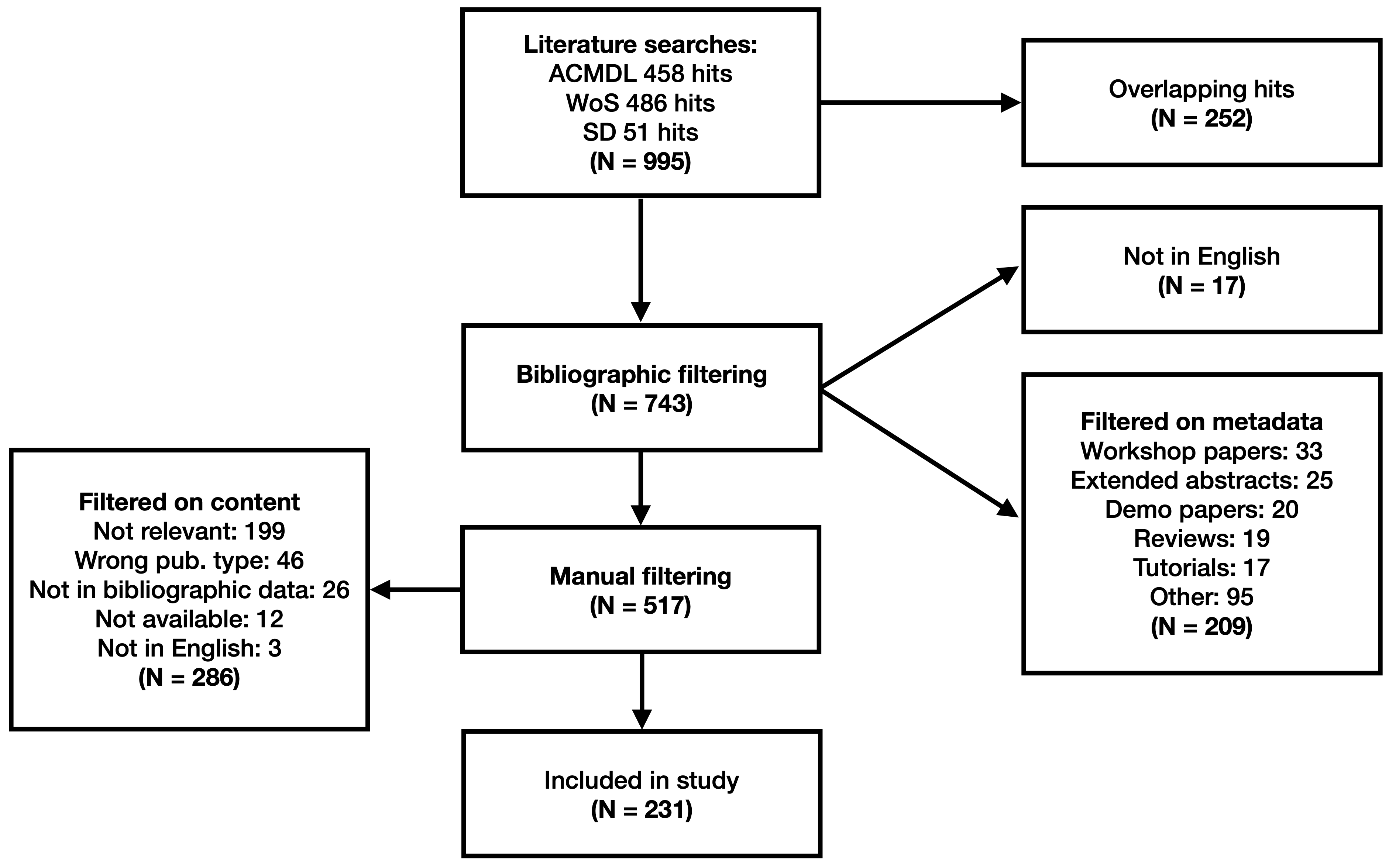}
  \caption{Article selection procedure overview.}
  \label{fig:flow}
\end{figure*}

\subsection{Literature Search}

Articles related to exploratory search can be found in multiple scientific conferences and journals covering various research disciplines. We selected three scientific repositories to conduct literature searches: ACM Digital Library (ACMDL), Web of Science (WoS, formerly Web of Knowledge) and Science Direct (SD). ACMDL provides access to $\mathord{\sim}$3 million articles related to computing through the ACM Guide to Computing Literature. WoS's core collection includes $\mathord{\sim}$80 million articles across all fields of science. SD contains $\mathord{\sim}$18 million articles from Elsevier publication venues across all fields of science.

For each scientific repository, we retrieved all publications where the exact phrase ``exploratory search'' appears in the bibliographic data (i.e.~title, abstract, keywords). 
WoS did not allow us to limit search terms to bibliographic data, so we searched using ``All fields''.
In general, we avoided full-text search as it returned too many articles that were not primarily about exploratory search. 
We limited our search to articles published between 2010--2021, inclusive. We did not limit search results on the basis of any other criteria, such as publication venue.

The literature search resulted in 995 search results (ACMDL: 458, WoS: 486, SD: 51). We identified duplicate articles between repositories based on DOIs (Digital Object Identifier) and, when the DOI was absent, article titles (duplicates identified via titles were manually checked). After duplicate search results were removed, 743 publications remained.

\subsection{Article Selection}

We imported the metadata data into Excel and manually screened articles to narrow down the set of papers to: 
\begin{enumerate*}[label=(\roman*)]
    \item peer-reviewed conference or journal articles, that were  
    \item written in English.
\end{enumerate*}
As metadata related to the type of publication was not always available, we inferred it using a combination of 
page count, 
article title (e.g.~{\em ``A review of ...''}), 
abstract ({\em ``In this survey, we ...''}) and 
publication venue ({\em ``Proceedings of the First International Workshop on ...''}).
This procedure excluded:
workshop papers (33),
extended abstracts (25), 
demo papers (20),
review articles (19),
tutorials (17), 
and other miscellaneous articles (95), including doctoral consortium papers, dissertations, and keynote speeches. 
%
%
We excluded a further 17 papers that were not written in English. 
To ensure this process was reliable, two authors independently marked papers for exclusion. Inter-rater reliability was assessed using Cohen's kappa 
($\kappa = 0.77$, 
$p < .00001$), indicating substantial agreement \cite{landis1977measurement}.
Disagreements were discussed between all authors and resolved by consensus. This procedure left 517 scientific articles.

\subsection{Validity Assessment}

In the validity assessment step, we screened out articles that were not substantively about exploratory search.
We excluded papers that were not related to information retrieval or information seeking (e.g.~the phrase is also used in optimization research), 
or where, despite appearing in bibliographic data, exploratory search is only peripherally mentioned in related work or discussion.

We downloaded the corresponding PDF files for all search results. 
Of the 517 papers from the article selection step, we excluded a further 286 papers for the following reasons: 
\begin{enumerate*}[label=(\roman*)]
    \item articles not about exploratory search (199),
    \item incorrect publication type, only discernible from the full text due to missing metadata (46), 
    \item ``exploratory search'' not found in title, abstract or keywords from WoS (26),
    \item unavailable online or inaccessible from our research institution (12), and
    \item body of the article was not written in English (3). 
\end{enumerate*}
%
%
Assessment was performed independently by all three authors. Inter-rater reliability 
was assessed using Fleiss' kappa, ($\kappa = 0.71$, $p < .00001$), indicating strong agreement. Any cases where the decision to exclude an article was unclear were discussed between the authors to reach a consensus.
After validity assessment, we were left with 231 papers, including journal articles, full-length and short conference papers, for use in our study. 

Following Webster and Watson, we took a concept-centric approach to our review where pre-defined units of analysis were checked and coded for in each paper as it was read \cite{webster2002analyzing}. These coded concepts were compiled into frequency tables, which we present next.

\section{Results}

In our results, we analyzed 
the interdisciplinarity of research in exploratory search (i.e.~{\em who conducts exploratory search research?}),
we enumerate retrieval systems, search domains and simulated work tasks ({\em what is the focus of exploratory search research?}),
and, lastly, detail the study characteristics of user studies based on simulated work tasks ({\em how do we conduct evaluation in exploratory search studies?}).
Several of the tables presented below are stratified by discipline (information science, information systems and human-computer interaction). However, for concision, this is only shown where per-discipline results differed substantially from the overall results.


\subsection{Interdisciplinarity} 

To characterize the interdisciplinary nature of research in exploratory search, 
we analyzed which journals and conferences each article was published in and 
how authors used article keywords to describe their work.

    \subsubsection{Publication venues}

\begin{table}[t]
\centering
\renewcommand*{\arraystretch}{1.1}
\resizebox{\columnwidth}{!}{%
\begin{tabular}{ l r }
\toprule
Publication venue                                                                   & Freq. \\ \midrule 
Conference on Human Information Interaction and Retrieval$^*$ (CHIIR)               & 15 \\  
Conference on Intelligent User Interfaces (IUI)                                     & 13 \\ 
Conference on Information and Knowledge Management (CIKM)                           & 10 \\ 
Information Processing \& Management (IPM)                                          & 9 \\ 
Journal of the Association for Information Science and Technology (JASIST)          & 9 \\ 
Conference on Research and Development in Information Retrieval (SIGIR)             & 9 \\ 
Association for Information Science and Technology (ASIS\&T)                        & 8 \\ 
Conference on Human Factors in Computing Systems (CHI)                              & 6 \\ 
Joint Conference on Digital Libraries (JCDL)                                        & 5 \\ 
The Web Conference (WWW)                                                            & 4 \\ \bottomrule
\end{tabular}
}
\caption{Most frequently occurring publication venues ($\ast$ = publication count for CHIIR includes both CHIIR (6) and IIiX (9)). }
\label{tab:venues}
\end{table}

    Table~\ref{tab:venues} shows the 10 most frequently occurring publication venues in our data set, 
    accounting for 38.1\% of articles. 
    These venues highlight the interdisciplinary nature and high standard of research in exploratory search, encompassing 
    information science (JASIST, ASIS\&T, JCDL), 
    information systems (IPM, CIKM, SIGIR, WWW) and 
    human-computer interaction (CHIIR, IUI, CHI). 
        
    Based on this finding, we categorised all 231 articles based on whether they were published in venues specialising in  
    information science (18.6\%), 
    information systems (35.5\%), 
    human-computer interaction (26.8\%), 
    and other (19.1\%). 
    The other category represents a wide variety of conferences and journals, 
    specialising in fields from artificial intelligence to educational technologies, 
    and multidisciplinary venues, such as PLOS One.

    \subsubsection{Author-defined keywords} 

Table~\ref{tab:keywords} shows the top-10 author-defined keywords for articles from information science, information systems and human-computer interaction. We lightly edited the keywords by making plurals singular and unified any differences between British and American spelling. 
While a majority of articles had author-defined keywords (91\%), 
the proportion varied between disciplines with information science being the lowest (81\%). 

The keyword distributions highlight the differences between disciplines that conduct research into exploratory search: 
there is a strong theme of information seeking in the information science articles  
(e.g.~{\em ``information seeking behavior''}, {\em ``search strategies''}, {\em ``information foraging''}).  
Whereas information systems focuses on technological aspects of search systems 
({\em ``information retrieval''}, {\em ``linked data''}, {\em ``semantic web''}) 
and human-computer interaction has numerous keywords related to information presentation 
({\em ``user interface''}, {\em ``information visualization''}, {\em ``visual analytics''}).
There are also commonalities between disciplines, with keywords found in multiple fields highlighting 
the dialogue-like back and forth between user and search system ({\em ``interactive information retrieval''}) 
and the central role of experimentation ({\em ``user study''}).


\begin{table*}[]
\centering
\renewcommand*{\arraystretch}{1.1}
\begin{tabular}{lrrlrrlrr}
\toprule
\multicolumn{3}{l}{Information Science} &
  \multicolumn{3}{l}{Information Systems} &
  \multicolumn{3}{l}{Human-Computer Interaction} \\
Keywords & Freq. & \% &
Keywords & Freq. & \% &
Keywords & Freq. & \% \\ 
\midrule
exploratory search                & 16 & 37.2 & exploratory search                & 54 & 65.9 & exploratory search        & 44 & 71.0 \\
information seeking               & 6  & 14.0 & information retrieval             & 7  & 8.5  & information seeking       & 6  & 9.7 \\
evaluation                        & 4  & 9.3  & faceted search                    & 5  & 6.1  & user study                & 6  & 9.7 \\
user study                        & 4  & 9.3  & search user interface             & 4  & 4.9  & user interface            & 4  & 6.5  \\
interactive ir                    & 4  & 9.3  & user study                        & 4  & 4.9  & sensemaking               & 4  & 6.5  \\
faceted search                    & 4  & 9.3  & linked data                       & 4  & 4.9  & information viz.          & 4  & 6.5  \\
information seeking behavior      & 3  & 7.0  & semantic web                      & 4  & 4.9  & collaborative search      & 3  & 5.6  \\
search strategies                 & 2  & 4.7  & information viz.                  & 4  & 4.9  & search behavior           & 3  & 5.6  \\
information foraging              & 2  & 4.7  & interactive ir                    & 4  & 4.9  & visual analytics          & 3  & 5.6  \\
serendipity                       & 2  & 4.7  & recommender system                & 3  & 3.7  & visualization             & 3  & 5.6 \\ 
\bottomrule
\end{tabular}%
\caption{Author-defined keywords from exploratory search articles in different disciplines.} 
\label{tab:keywords}
\end{table*}

\subsection{Study Settings}

    
    Of the 231 papers included in our study, 172 included one or more user studies. We used this subset of papers to understand what empirical studies of exploratory search investigate in terms of retrieval systems, search domains and simulated work tasks.
    
    \subsubsection{Retrieval Systems}
    
    Table~\ref{tab:systems} shows the named retrieval systems from the 172 articles that included user studies. We further categorised each system as {\em experimental} and {\em production} systems. Experimental systems tend to be prototypes with articles often evaluating new features, whereas production systems are available on the open Internet, such as commercial search engines (e.g.~Bing) and digital libraries (Sowiport). We did not count baseline systems unless they were a named retrieval system (i.e.~``list-based interfaces'' or baselines comprised of the experimental system without specific features were not included). Systems that appeared only once in the data set were collapsed into a ``miscellaneous'' subcategory.
    
    A majority of exploratory search user studies are based on the evaluation of experimental search systems (77\%) 
    with the most frequently occurring systems being Coagmento \cite{shah2010coagmento} and SciNet \cite{glowacka2013directing}. These studies tended to evaluate novel interface components that support users performing exploratory search (e.g.~\cite{qvarfordt2013looking}). Whereas, studies based on production systems focused primarily on investigating exploratory search behavior. For example, \cite{athukorala2016exploratory} performed a comparative study of lookup and exploratory search behavior 
    using Google Scholar. 

\begin{table}[]
\centering
\renewcommand*{\arraystretch}{1.1}
\resizebox{0.85\columnwidth}{!}{%
\begin{tabular}{llrr}
\toprule
           & System Name       & Freq. & \% \\ \midrule
Experimental   & Coagmento         & 8     & 4.6 \\
           & SciNet            & 6     & 3.4 \\
           & Discovery Hub     & 4     & 2.3 \\
           & Cruise            & 3     & 1.7 \\
           & CollabSearch      & 2     & 1.1 \\
           & ConceptCloud      & 2     & 1.1 \\
           & Paths             & 2     & 1.1 \\
           & Querium           & 2     & 1.1 \\
           & uRank             & 2     & 1.1 \\
           & Misc.             & 103    & 59.2 \\ \midrule
Production & User-selected Search Engine & 15  & 8.6  \\
           & Bing              & 5     & 2.9 \\
           & Google            & 5     & 2.9 \\
           & Sowiport          & 3     & 1.7 \\
           & Google Scholar    & 2     & 1.1 \\
           & Misc.             & 10    & 5.7 \\ \midrule
{\bf Total}      &             & {\bf 174}$^\ast$ &   \\ 
\bottomrule
\end{tabular}%
}
\caption{Named retrieval systems in our data set ($\ast$ = total is greater than the number of papers with user studies (N=172) as two papers compared two of the listed systems). }
\label{tab:systems}
\end{table}

    \subsubsection{Search Domains} 

\begin{table*}[]
\centering
\renewcommand*{\arraystretch}{1.1}
\begin{tabular}{llrrrrrrrr}
\toprule
 & & \multicolumn{2}{c}{Info.~Science} & \multicolumn{2}{c}{Info.~Systems} & \multicolumn{2}{c}{HCI} & \multicolumn{2}{c}{All (incl.~other)} \\
Domain                & Subdomain                           & Freq. & \%  & Freq. & \%    & Freq. & \%    & Freq. & \% \\ \midrule
Data                  & DBpedia                             &    &        & 6  & 8.3      &    &          & 8   &  3.8         \\
                      & Experimental Results                &    &        & 1  & 1.4      & 2  & 3.2      & 3   &  1.4         \\
                      & Search Logs                         &    &        & 2  & 2.8      &    &          & 2   &  0.9         \\
                      & Academics                           &    &        & 2  & 2.8      &    &          & 2   &  0.9         \\
                      & Misc.                               & 1  & 2.2    & 3  & 4.2      & 4  & 6.3      & 7   &  3.3         \\ \midrule
Documents             & News                                &    &        & 1  & 1.4      & 1  & 1.6      & 6   &  2.8         \\
                      & Digital Heritage                    & 3  & 6.5    &    &          & 1  & 1.6      & 4   &  1.9         \\
                      & Health                              &    &        & 2  & 2.8      & 1  & 1.6      & 4   &  1.9         \\
                      & Misc.                               & 4  & 8.7    & 7  & 9.7      & 7  & 11.1     & 23  &  10.8        \\ \midrule
Multimedia            & Images                              & 1  & 2.2    & 4  & 5.6      & 2  & 3.2      & 9   &  4.2         \\
                      & Various$^{\ast}$                    & 1  & 2.2    & 1  & 1.4      & 2  & 3.2      & 5   &  2.4         \\
                      & Video                               & 2  & 4.3    & 1  & 1.4      &    &          & 3   &  1.4         \\
                      & Audio                               & 1  & 2.2    &    &          &    &          & 1   &  0.5         \\ \midrule
Scientific            & Computer Science                    & 5  & 10.9   & 12 & 16.7     & 17 & 27.0     & 36  &  17.0        \\
Literature            & Multiple Disciplines$^{\ast\ast}$   & 1  & 2.2    &    &          & 1  & 1.6      & 3   &  1.4         \\
                      & Social Sciences                     & 3  & 6.5    &    &          &    &          & 3   &  1.4         \\
                      & Life Sciences                       &    &        &    &          & 1  & 1.6      & 2   &  0.9         \\ \midrule
Web                   & Technology                          & 6  & 13.0   & 4  & 5.6      & 4  & 6.3      & 16  &  7.5         \\
                      & Environment                         & 3  & 6.5    & 4  & 5.6      & 4  & 6.3      & 12  &  5.7         \\
                      & Health                              & 1  & 2.2    & 3  & 4.2      & 3  & 4.8      & 10  &  4.7         \\
                      & Travel                              & 2  & 4.3    & 3  & 4.2      & 2  & 3.2      & 8   &  3.8         \\
                      & Education                           & 1  & 2.2    & 1  & 1.4      & 1  & 1.6      & 7   &  3.3         \\
                      & History                             & 2  & 4.3    & 4  & 5.6      &    &          & 7   &  3.3         \\
                      & News                                & 4  & 8.7    &    &          & 1  & 1.6      & 5   &  2.4         \\
                      & Misc.                               & 5  & 10.9   & 11 & 15.2     & 9  & 14.3     & 26  &  12.3        \\ \midrule
{\bf Total}           &                                     & {\bf 46} &  & {\bf 72} &    & {\bf 63} &    & {\bf 212$^{\ast\ast\ast}$} & \\ \bottomrule
\end{tabular}%
\caption{Overview of domains used in exploratory search articles with user studies stratified by discipline (N=172). All column includes studies outside of information science, information systems and human-computer interaction. 
($\ast$ = containing multiple multimedia data types, 
$\ast\ast$ = domain decided based on participant's background, 
$\ast\ast$$\ast$ = many user studies covered multiple subdomains).}
\label{tab:domains}
\end{table*}

    Table~\ref{tab:domains} shows the domains used in exploratory search user studies stratified by discipline. We grouped our coding of domains into 5 categories: {\em data}, {\em documents}, {\em multimedia}, {\em scientific literature} and {\em web}. 
    Data included structured data, such as knowledge graphs (most commonly DBpedia \cite{lehmann2015dbpedia}) and log data.
    Documents included unstructured corpus data, such as news articles.
    Multimedia included audio, video and image data, with ``various'' indicating multiple media types.
    Scientific literature was coded by discipline, with an additional ``multiple disciplines'' subdomain for search tasks where the participant's background was relevant.
    Lastly, web included different domains of web search, such as travel and politics, following articles' descriptions. In all domains, we included a ``miscellaneous'' category to group infrequently occurring subdomains. Each article could include multiple domains or multiple subdomains within a given domain.

    All domain categories were present in all three disciplines. 
    The most common domains overall were web (42.9\%) and scientific literature (20.7\%). The least common domain was multimedia (8.5\%).
    We note that user studies could include multiple subdomains. 
    For example, web search covers 42.9\% of {\em subdomains} listed in Table~\ref{tab:domains}, but only 30.8\% of {\em articles}.
    Studies in the data domain were primarily from information systems where exploratory search is studied in connection to semantic web (e.g.~\cite{thakker2013assisting}).
    Interestingly, a majority of the studies in scientific literature domain were related to Computer Science (accounting for 17.0\% of subdomains and 20.9\% of articles). This was driven by studies in information systems and human-computer interaction, which likely had access to Computer Science students who need to perform learning tasks in the completion of their dissertations. For example, Sciascio et al.~investigated ideas from social search to enhance the exploratory search of scientific literature for Master's and doctoral students \cite{di2018study}.
    



    \subsubsection{Tasks} 

\begin{table}[]
\centering
\renewcommand*{\arraystretch}{1.1}
\resizebox{\columnwidth}{!}{%
\begin{tabular}{llrr}
\toprule
                        & Task Type                & Freq.  & \%    \\ \midrule
Exploratory Search      & Knowledge Acquisition    & 62     & 33.2     \\
                        & Planning                 & 50     & 26.7     \\
                        & Comparison               & 7      & 3.7     \\ \midrule
Lookup Search           & -                        & 11     & 5.9     \\ \midrule
No Simulated Work Task  & Usability Assessment     & 30     & 16.0     \\
                        & Search Results Assessment & 16     & 8.6     \\
                        & Log Data Analysis        & 11     & 5.9     \\ \midrule
{\bf Total}             &                          & {\bf 187$^\ast$} &   \\ 
\bottomrule
\end{tabular}%
}
\caption{Types of task performed in exploratory search articles with user studies (N=172).\\
($\ast$ = many articles contained user studies that involved multiple task types).}
\label{tab:tasktype}
\end{table}

Table~\ref{tab:tasktype} shows three categories of search task performed during exploratory search user studies: exploratory search, lookup search and no simulated work task. We note that several articles contained multiple user studies. A majority of studies (63.6\%) included simulated work tasks \cite{borlund2003iir} for exploratory search. Following \cite{athukorala2016exploratory}, we further subdivided exploratory search tasks into knowledge acquisition (learning tasks with open-ended search goals \cite{wu2012grannies}), planning (investigation tasks that are focused on producing a general overview of a topic \cite{wu2012grannies}) and comparison (gathering information about two or more topics to identify similarities and differences \cite{liu2010personalizing}). Very few studies contrasted exploratory search with lookup search tasks (5.9\%). We categorised studies with no simulated work task into usability assessment, search results assessment and log data analysis (30.5\%).

Exploratory search studies with simulated work tasks feature a reasonably balanced combination of learning and investigation tasks.
Several studies contrast exploratory and lookup search to identify differences in user search behavior.
Lastly, studies with no simulated work task tend to focus on usability assessment, for example, in interface evaluation and co-design activities.




\subsection{Evaluation Methodologies}

To understand the experimental characteristics of exploratory search user studies, we focused on the 115 articles that included 124 studies with simulated work tasks. 
We investigated the experimental design, choice of study participants and methods of data collection.


    \subsubsection{Experimental Design}

\begin{table}[t]
\centering
\renewcommand*{\arraystretch}{1.1}
\begin{tabular}{ lrr }
\toprule
Experimental Design & Freq. & \% \\ \midrule 
Within-subject      & 62 & 50.0 \\
Observational       & 31 & 25.0 \\
Between-subject     & 23 & 18.5 \\
Mixed               & 8  & 6.5  \\ 
\bottomrule
\end{tabular}
\caption{Experiment design in studies with simulated work tasks (N=124).}
\label{tab:experimentaldesign}
\end{table}
    

    In user studies where participants performed simulated work tasks, sample sizes ranged from 2 to 381. The median sample size was 24 and the mean sample size was 35.7 (95\% CI [28.0, 43.5]). 
    A majority of exploratory search user studies were experiments (75\%), i.e.~non-observational studies (see Table~\ref{tab:experimentaldesign}). The most common experimental design was within-subject, which had lower average sample sizes (median = 20, mean = 32.1 (95\% CI [19.6, 44.6])) than between-subject studies (median = 41.5, mean = 50.0 (95\% CI [36.4, 63.5])). In general, there is a lack of large-scale studies in exploratory search: we found 5 user studies with sample sizes greater than 100 participants. The largest of these studies, however, pooled observations from multiple past experiments \cite{hendahewa2017evaluating} or were focused on evaluating a questionnaire in the context of exploratory search \cite{o2013examining}.


    
    \subsubsection{Choice of Participants}

    We extracted details about study participants from the 124 studies that used simulated work tasks (see Table~\ref{tab:participants}). 
    As participant descriptions are informal and often ambiguous (e.g.~``postgraduate'' and ``doctoral student'' may be synonyms in some countries, but not others), we categorized participants into {\em academics} and {\em non-academics}, but otherwise did not want to over-interpret their descriptions. 
    

    In a majority of studies, participants either included academics (83.9\%) or were exclusively composed of academics (75.0\%). We additionally extracted disciplines from the descriptions of academic participants. The two most common disciplines were {\em Unstated} (38.9\%) and {\em Computer Science} (30.6\%) (data not shown). The most common types of non-academic participant were without any description (10.5\%) or described as the general public (8.9\%). It is possible that non-academic participants had academic backgrounds. 
    %


\begin{table}[]
\centering
\renewcommand*{\arraystretch}{1.1}
\resizebox{\columnwidth}{!}{%
\begin{tabular}{llrr}
\toprule
Category      & Type of Participant           & Freq.   & \%    \\ \midrule
Academics     & Postgraduates                 & 44      & 35.5     \\
              & Undergraduates                & 39      & 31.5     \\
              & Students                      & 24      & 19.4     \\
              & Doctoral Students/Researchers & 22      & 17.7     \\
              & Academic Staff                & 18      & 14.5     \\
              & Postdoctoral Researchers      & 11      & 8.9      \\
              & From a University$^\ast$      & 11       & 8.9     \\ \midrule 
Non-academics & Non-descript                  & 13      & 10.5     \\ 
              & General Public                & 11      & 8.9      \\
              & Domain Experts/Employees/Teachers/Authors & 6 & 4.8  \\
              & High School Students          & 1       & 0.8      \\
              & Mechanical Turk               & 1       & 0.8      \\ 
\bottomrule
\end{tabular}%
}
\caption{Participants in studies with simulated work tasks (N=124).}
\label{tab:participants}
\end{table}

    \subsubsection{Data Collection} 

Table~\ref{tab:data} shows the data collected in the 124 studies that included simulated work tasks. 
We categorised each measure as  
{\em interaction}, {\em performance} or {\em usability} \cite{kelly2009methods}. 
Interactions are observations used to understand search behavior (e.g.~search queries, think-aloud, etc.). We collectively refer to interface-related interactions other than search queries, clicks and bookmarks as {\em interface interactions} (i.e.~interactions specific to a given retrieval system).
Performance measures are used to judge the success of a simulated work task (e.g.~task time).
Usability included questionnaires and interviews to collect information about users' experiences and preferences. 

The most common interaction measures were easy to collect events, such as queries, clicks and bookmarks, with few studies capturing qualitative indicators of search behavior (think-aloud, eye tracking, etc.) that would require additional coding or more complex analysis.
Similarly, the most frequently occurring performance measures were those that can be quantified, such as task time, free-form summaries (graded by a domain expert) and relevant documents (retrieved or bookmarked documents classified by a domain expert or the participant). While there have been attempts to develop novel performance indicators, such as concept maps \cite{egusa2010using, egusa2014concept}, to evaluate knowledge acquisition, such methods have not been widely-adopted by the research community.
Lastly, very few studies used standardized questionnaires, but asked similar questions in self-defined (unvalidated) questionnaires. Furthermore, many of the studies that used standardized questionnaires tended to use more than one. For example, \cite{medlar2021query} used both SUS \cite{brooke1996sus} and ResQue \cite{pu2011user} to investigate usability and user experience, respectively.

\begin{table*}[]
\centering
\renewcommand*{\arraystretch}{1.1}
\begin{tabular}{lrrlrrlrr}
\toprule
Interaction            & Freq. & \% & Performance        & Freq. & \% & Usability                     & Freq. & \% \\ \midrule
Search Queries         & 61    & 49.2 & Task Time          & 36    & 29.0 & Self-defined Questionnaires   & 53  & 42.7   \\
Interface Interactions & 61    & 49.2 & Free-form Summary  & 23    & 18.5 & Interviews                    & 41  & 33.1   \\
Clicked Documents      & 53    & 42.7 & Relevant Documents & 18    & 14.5 & SUS (Usability)               & 12  & 9.7    \\
Bookmarks              & 42    & 33.9 & Question Answers   & 12    & 9.7  & ResQue (UX)                   & 11  & 8.9    \\
Think-Aloud            & 16    & 12.9 & Relevant Snippets  & 5     & 4.0  & NASA TLX (Workload)           & 7   & 5.6    \\
Screen Capture         & 14    & 11.3 & Concept Maps       & 3     & 2.4  & UES (Engagement)              & 3   & 2.4    \\
Snippets               & 12    & 9.7  & Research Question  & 1     & 0.8  & USE (Usability)               & 1   & 0.8    \\
Eye Tracking           & 10    & 8.1  & Essays             & 1     & 0.8  & PANAS (Affect)                & 1   & 0.8    \\
Chats                  & 8     & 6.4  &                    &       &      & TAM (Acceptance)              & 1   & 0.8    \\
Video Recording        & 3     & 2.4  &                    &       &      &                               &     &        \\
User Notes             & 3     & 2.4  &                    &       &      &                               &     &        \\
Audio Recording        & 1     & 0.8  &                    &       &      &                               &     &        \\ \bottomrule
\end{tabular}%
\caption{Data collected in user studies with simulated work tasks (N=124).}
\label{tab:data}
\end{table*}

\section{Discussion}

This review has presented a meta-analysis of empirical studies in exploratory search. We were motivated by the lack of recent reviews of exploratory search and chose the period of 2010-2021 to continue on from White and Roth's review \cite{white2009exploratory}. Furthermore, we wanted to focus on empirical research to enumerate what is studied (systems, domains and tasks) and how it is studied (experimental designs, participants, data collection). The following discussion is a critical analysis of our findings and how they should influence future research in exploratory search.

\subsection{Interdisciplinarity}

Our review findings confirm the interdisciplinary nature of exploratory search with a majority of articles published in journals and conferences dedicated to information science, information systems and human-computer interaction (80.9\%). Our findings also highlight the quality of that research: the ten most frequently occurring publication venues account for a large proportion of our data set (38.1\%) and include the top-ranked venues in each discipline. 
While 
our analysis of author-defined keywords suggests that the core focus of articles is generally discipline-specific, we believe this is related to how research is positioned for a given audience as 
\begin{enumerate*}[label=(\roman*)]
    \item the most prolific authors in our data set published articles in more than one discipline and 
    \item outside of search domains, we found very little distinction between disciplines at the level of granularity presented in this review.
\end{enumerate*}

We were surprised to find so few articles published in venues dedicated to disciplines that utilize search, such as medicine (e.g.~\cite{pang2016designing}) and education (e.g.~\cite{cortinovis2019supporting}), in our data set. 
Future studies should seek to widen the scope of domains in which exploratory search is applied. This could lead to a more comprehensive understanding of how exploratory search manifests in different search contexts and identify novel challenges. Furthermore, such studies should be published in venues that correspond to the specific domain under study, e.g.~studies on legal search in law journals, to ensure a high degree of rigor during the peer-review process.


\subsection{Study Settings}

We identified significantly more empirical studies based on experimental retrieval systems (77\%) than production systems, such as commercial search engines (23\%). This shows that a majority of studies were focused on the evaluation of new technologies and interfaces to support searchers rather than investigating users' search processes with publicly available search systems. Furthermore, there appears to be very little software reuse between studies, with SciNet \cite{glowacka2013directing} and Coagmento \cite{shah2010coagmento} being the main exceptions. However, SciNet is not publicly available and Coagmento is intended for collaborative search and not specifically exploratory search. 
Future work in exploratory search should have a greater emphasis on software reuse and reproducibility. The source code of retrieval systems should be open source and, where possible, log data from anonymized search sessions made available for conducting meta-analyses (e.g.~\cite{medlar2018consistent}).

A majority of empirical studies used simulated work tasks to situate users in exploratory search scenarios (63.6\%). While we identified similar numbers of simulated work tasks for learning and investigation, very few studies included lookup search tasks. Therefore, few studies could attempt to create search systems that could adapt to whether users were conducting lookup or exploratory search \cite{athukorala2016exploratory, athukorala2016beyond, medlar2017towards}. 
While empirical studies have been conducted in numerous search domains, there is a clear bias toward web search (43.0\%) and scientific literature search (20.7\%). Furthermore, some domains were either only found in a single discipline (the data domain was almost exclusively found in articles from information systems) or concentrated in a single subcategory (81.8\% of studies based on scientific literature search were limited to Computer Science articles). 
Future work on exploratory search must broaden the scope of research beyond current search domains to enable the study of more diverse user populations, search tasks and information modalities (e.g.~\cite{kropotov2021exploratory, liu2022rogue}).



%


\subsection{Evaluation Methodology}

From a methodological perspective, our review highlights many of the issues with empirical studies in exploratory search.
Sample sizes are generally low, with calculations for statistical power either rarely reported or not conducted. Either way, within-subject and between-subject experimental designs likely had similar statistical power with median sample sizes of 20 and 41.5, respectively. In general, studies used convenience samples of participants from the authors' universities. Indeed, 75\% of studies were based exclusively on students and academics. This lack of engagement with participants outside of academia is concerning and shows that our understanding of exploratory search is limited to WEIRD populations \cite{henrich2010weirdest}. Only a small proportion of studies included members of the general public and we only identified a single study that used a crowdsourcing platform \cite{cui2014evaluation}.
Future research on exploratory search should focus on recruiting study participants from the broader population. In particular, experiments should be designed for crowdsourcing platforms (i.e.~including control mechanisms to ensure reliability) which have been shown to be more representative of the general population than laboratory studies \cite{difallah2018demographics} and will allow for larger sample sizes.


While empirical studies of exploratory search used numerous methods in evaluation, they tended to be based on quantitative interaction and performance variables that are easily measured, such as search queries, clicks and task time. 
In general, there was little consensus on what variables were used to measure interaction behavior, whereas performance measures tended to be task-specific, e.g.~using free-form summaries to simulate essay planning tasks.
Moreover, qualitative research methods, such as think-aloud, were rarely used despite being one of the few techniques available to investigate the cognitive and interpretive aspects of users' search processes. While interviews were used in many studies, they suffer from being retrospective and are unable to capture the fleeting thoughts and iterative nature of exploratory search. Anecdotally, we observed that interviews were rarely coded, but mined for quotations to support quantitative results. Lastly, while usability was measured using questionnaires, these were generally non-validated, with standardised questionnaires, such as SUS and ResQue, rarely used. 
Future research in exploratory search should make better use of qualitative research methods to understand users' search processes and validated questionnaires should be preferred to ensure that results are robust and comparable between studies.


\subsection{Limitations}

The aim of this article was to provide a comprehensive review of empirical studies in exploratory search. The review process was systematic and, therefore, reproducible. However, our review also has several limitations. 
First, our review could only provide an overview of exploratory search as the process of conducting a systematic review necessarily leads to the nuances of individual studies to be lost. While our review did not elaborate on the details of individual studies, we believe that it can be used as a starting point for more focused reviews in the future.
Second, despite exploratory search being recognized as an important topic in information retrieval and search, it is a relatively niche area of research. Prolific research groups, such as the authors of SciNet, for example, have the potential to skew the distributions reported in our results section. 
Third, many of the articles that were not categorised as information science, information systems or human-computers interaction added little value to the review. However, we decided to keep these articles to avoid biasing results to only top venues.
Fourth, our study excludes papers that define themselves as searching as learning. While several of the articles in our review used both terms, our study design was such that it required papers to be explicitly related to exploratory search. We similarly omitted papers positioned as interactive information retrieval and cognitive search that could have fit the definition of exploratory search.
Fifth, we may have omitted important papers due to the selection of databases. We selected databases based on our understanding of the type of publication venue that would accept work in exploratory search and, therefore, believe that the number of omissions is likely to be small in number and would not have substantially impacted our overall conclusions.

\bibliographystyle{ACM-Reference-Format}
\bibliography{sample-manuscript}

\end{document}